\documentclass{elsart}

\usepackage{amssymb}

\begin{document}

\begin{frontmatter}



\title{Energy density in general relativity: a possible role for cosmological constant}

\author[label1,label2]{Saibal Ray} \author[label3]{Sumana Bhadra}
\address[label1]{Department of Physics, Barasat Government College,
Barasat 700 124, North 24 Parganas, West Bengal, India}
\address[label2]{Inter-University Centre for Astronomy and Astrophysics, Post Box
4, Pune 411 007, India}
\address[label3]{Balichak Girls' High School, Balichak 721 124,
West Midnapur, West Bengal, India}



\begin{abstract}
We consider a static spherically symmetric charged anisotropic
fluid source of radius $\sim 10^{-16}$ cm by introducing a
variable $\Lambda$ dependent on the radial coordinate $r$ under
general relativity. From the solution sets a possible role of the
cosmological constant is investigated which indicates the
dependency of energy density on it.
\end{abstract}


\end{frontmatter}

\section{Introduction}
\label{}



The structure of electron was, for a long time, an intrigue problem to
the researchers. Many scientists, like Lorentz [1] and even Einstein
[2] tried to solve the problem to show that the electron mass is a
electromagnetic field dependent quantity (for a detail account see the
references [3] and [4]). Lateron, under general relativity some
models have been proposed by different authors describing extented
electron with its mass entirely of electromagnetic origin [5 - 8]. Recently,
based on the experimental upper limits on the size of the electron as
$\sim 10^{-16}$ cm [9] it is argued by Cooperstock and Rosen [10],
Bonnor and Cooperstock [11] and Herrera and Varela [12] that in the
framework of general theory of relativity the electron-like
spherically symmetric charged distribution of matter must contain some
negative mass density. Motivated by these results with historical and
heuristic values we would like to explore a possible role of
cosmological constant on the energy density of electron when it is
modeled as a variable dependent on the radial coordinate $r$ of the
charged spherical matter distribution.

The basic logic for considering variability of so called cosmological
constant, which was introduced by Einstein in 1917 to obtain a static
 cosmological model, is related to the observational
evidence of high redshift Type Ia supernovae [13, 14] for a small
decreasing value of cosmological constant ($\Lambda_{present}\leq
10^{-56} cm^{-2}$) at the present epoch. This indicates that
instead of a strict constant the $\Lambda$ could be a function of space
and time coordinates. If the role of time-dependent $\Lambda$ is
prominent in the cosmological realm, then space-dependent $\Lambda$ has an
expected effect in the astrophysical context. It is, therefore, argued
by Narlikar et al. [15] that the space-dependence of $\Lambda$ cannot be
ignored in relation to the nature of local massive objects like
galaxies. Our aim, however, to see if there is any effect of space-dependent
$\Lambda$ on the energy density of the classical electron.
This is because cosmological constant is thought to be related to the quantum
fluctuations as evident from the theoretical works by Zel'dovich [16].
Moreover, it is believed through indirect evidences that 65 \% of the
contents of the universe is to be in the form of the energy of vacuum
[17]. Thus, the energy  density of vacuum due to quantum fluctuation
might have, in our opinion, some underlying relation to the energy
density of Lorentz's extended electron [1] under general relativistic
treatment.

In the present letter we have tried to find out, through some specific
case studies, that energy density of classical electron is related to the
variable cosmological constant and the gravitational mass of the electron is
entirely dependent on the electromagnetic field alone.

\section{The field equations}
 To carry out the investigation we have considered the
 Einstein-Maxwell field equations for the case of anisotropic charged
fluid distribution (in relativistic units $G = c = 1$) which are given by
\begin{equation}
{G^{i}}_{j} = {R^{i}}_{j} - {{g^{i}}_{j}} R/2 = -8\pi [{{T^{i}}_{j}}^{(m)}+{{T^{i}}_{j}}^{(em)}+ {{T^{i}}_{j}}^{(vac)} ],
\end{equation}
\begin{equation}
{[{(-g)}^{1/2}F^{ij}], }_{j}= 4\pi J^{i}{(-g)}^{1/2},
\end{equation}
\begin{equation}
F_{[ij, k]}= 0
\end{equation}
where ${F^{ij}}$ is the electromagnetic field tensor and ${J^{i}}$,
 current four vector which is equivalent to ${J^{i}}= \sigma {u^{i}}$,
 $\sigma $ being the charge density and $u^{i}$ is the four-velocity
 of the matter satisfying the relation $u_{i}u^{i} = 1$.

The matter, electromagnetic and vacuum energy-momentum tensors are, respectively given by
\begin{equation}
{{T^{i}}_{j}}^{(m)} = (\rho + p_{\perp}) u^{i}u_{j} - p_{\perp} { g^{i}}_{j} +(p_{\perp} -p_{r}){\eta}^{i}{\eta}_{j},
\end{equation}
\begin{equation}
{{T^{i}}_{j}}^{(em)}= - [ F_{jk}F^{ik} - {g^{i}}_{j}F_{kl}F^{kl}/4]/4\pi,
\end{equation}
\begin{equation}
 {{T^{i}}_{j}}^{(vac)}= {{g^i}_j}\Lambda(r)/8\pi
\end{equation}
where  $\rho$,  $p_r$ and  $p_{\perp}$  are the proper energy density,
radial and tangential pressures respectively and also $\eta_i$ is the
unit spacelike vector on which the condition to be imposed is
${\eta}_{i}{\eta}^{i} = - 1$. Here $p_r$ is the pressure in the direction
of $\eta_i$ whereas $p_{\perp}$ is the pressure on the two-space
orthogonal to $\eta_i$.

Now, for the spherically symmetric metric
\begin{equation}
ds^2 = e^{\nu(r)}dt^2 - e^{\lambda(r)}dr^2 - r^2(d{\theta^2} +
 sin^2{\theta}d{\phi}^2),
\end{equation}
the Einstein-Maxwell field equations (1) - (6) corresponding to
anisotropic charged fluid with spatially varying cosmological
constant i.e. $\Lambda = \Lambda(r)$, are given by
\begin{equation}
e^{-\lambda}(\lambda^\prime/r - 1/r^2) + 1/r^2 =
8\pi T^0_0 = 8\pi{\tilde {\rho}} + E^2,
\end{equation}
\begin{equation}
e^{-\lambda}({\nu}^{\prime}/r + 1/r^2) - 1/r^2
= - 8\pi{{T^1}_1} = 8\pi \tilde{ p_r} - E^2,
\end{equation}
$$\hspace{-2in}e^{-\lambda}[{\nu}^{{\prime}{\prime}}/2 + {{\nu}^{\prime}}^2/4 -
{{\nu}^{\prime}}{{\lambda}^{\prime}}/4 + ({\nu}^{\prime} -
{\lambda}^{\prime})/2r]$$
\begin{equation}
~~~~~~~~~~~~~~~~~~= - 8\pi{{T^2}_2} = - 8\pi{{T^3}_3}
= 8\pi \tilde {p_{\perp}} + E^2,
\end{equation}
\begin{equation}
[r^2 E]^{\prime} = 4\pi r^2 {\sigma} e^{{\lambda}/2},
\end{equation}
where $E$, the intensity of electric field, is defined as
$ E = -e^{-(\nu + \lambda)/2}{\phi}^{\prime}$ and can equivalently be
expressed, from equation (11), as
\begin{equation}
E = \frac{1}{r^2}\int^{r}_{0} 4\pi r^2 {\sigma} e^{\lambda/2} dr.
\end{equation}
Here prime denotes derivative with respect to the radial coordinate $r$
only.

 In the above equations (8) - (10) we have considered that
\begin{equation}
\tilde {\rho}  = \rho + \Lambda(r)/8\pi,
\end{equation}
\begin{equation}
{\tilde p}_r = p_r - \Lambda(r)/8\pi,
\end{equation}
\begin{equation}
{\tilde p}_{\perp} = p_{\perp} - \Lambda(r)/8\pi,
\end{equation}
where ${\tilde {\rho}}$, ${{\tilde p}_r}$ and ${{\tilde p}_{\perp}}$ are
the effective energy density, radial and tangential
pressures respectively.

 The equation of continuity ${{T^i}_j};i = 0$, is given by
\begin{equation}
\frac{d{p_r}}{dr} - \frac{1}{8\pi}\frac{d\Lambda(r)}{dr} + \frac{1}{2}(\rho+p_r){\nu^\prime}
=\frac{1}{8\pi r^4}\frac{dq^2}{dr} + \frac{2(p_{\perp}-p_r)}{r}
\end{equation}
where $q$ is the charge on the spherical system.

We assume the relation between the radial and tangential pressures [12] as
\begin{equation}
 p_{\perp} - p_r = \alpha q^2r^2,
\end{equation}
where $\alpha$ is a constant.

Hence, by use of equations (14) and (17), the equation (16) reduces to
\begin{equation}
\frac{d \tilde {p_r}}{dr}+\frac{1}{2}( \tilde {\rho} +
 {{\tilde p}_r}){{\nu}^{\prime}}
 =\frac{1}{8\pi r^4}\frac{dq^2}{dr}+2{\alpha}q^{2}r.
\end{equation}
Again, equation (8) can be expressed in the following form as
\begin{equation}
 e^{-\lambda}= 1 - 2M/r,
\end{equation}
where the active gravitational mass, $M$, is given by
\begin{equation}
M = 4{\pi}{\int_0^r}\left[\tilde {\rho} +\frac{E^2}{8{\pi}}\right]r^2dr.
\end{equation}

\section{The solutions}
\subsection{Model for $\rho +p_r = 0$}
 Let us now solve the equation (18) under the assumption between the
stress-energy tensors as ${T^1}_1 = {T^0}_0$, which implies that
\begin{equation}
\tilde{\rho} + \tilde{ p_r} = \rho + p_r = 0.
\end{equation}
Also to make the equation (12) integrable we assume that
\begin{equation}
\sigma = {\sigma}_0 e^{-\lambda/2},
\end{equation}
where $\sigma_0$ is the charge density at $r = 0$ of the spherical
distribution, i.e. the central density of charge.

Hence, using condition (22) in equation (12), we get for the expression of electric charge and intensity of the electric field as
\begin{equation}
q = Er^2 = \frac{4}{3}\pi{\sigma}_0r^3.
\end{equation}
With the help of equations (21) and (23), the equation (18) reduces to
\begin{equation}
\frac{d \tilde {p_r}}{dr}=\frac{4}{3}\pi{\sigma_0}^2r+2{\alpha} q^2r,
\end{equation}
Thus the  solution set is given by
\begin{equation}
e^{-\lambda}=e^{\nu}=1-\frac{16}{45}{\pi}^2{\sigma_0}^2r^2(5a^2-2r^2)-\frac{8}{15}\pi\alpha q^2r^2(5a^2-3r^2),
\end{equation}
\begin{equation}
p_{r}= - (\alpha q^2 +\frac{2}{3}{\pi}{{{\sigma}^2}_{0}})(a^{2}-r^{2})
+ \frac{\Lambda(r)}{8\pi},
\end{equation}
\begin{equation}
p_{\perp}= \alpha q^2r^2 - (\alpha q^2 + \frac{2}{3}\pi{\sigma}^2_0)(a^2-r^2)+\frac{\Lambda(r)}{8\pi},
\end{equation}
\begin{equation}
\rho(r)=(\alpha q^2 + \frac{2}{3}\pi\sigma^2_0)(a^2-r^2)- \frac{\Lambda(r)}{8\pi}.
\end{equation}
The active gravitational mass which is defined in the equation (20),
then, by virtue of the equations (23) and (28), takes the form as
\begin{equation}
M(r) = \frac{8}{135}{\pi}^2 {\sigma_{0}}^2 r^3 [8 \pi \alpha a^6 ( 5 a^2 - 3r^2 ) + 3(
5a^2 - 2r^2 )].
\end{equation}
Thus, the metric potentials $\lambda$ and $ \nu$ are given by
\begin{equation}
e^{-\lambda} = e^ {\nu} = 1 - \frac{2M(r)}{r}.
\end{equation}
The total effective gravitational mass can be obtained, after smoothly
matching of the interior solution to the exterior Reissner-Nordstr{\"o}m
solution on the boundary, as
\begin{equation}
m = M(a) + \frac{q^2(a)}{2a} = \frac{64}{45} {\pi}^2 {\sigma_{0}}^2 a^5 (1 + \frac{2}{3}\pi \alpha a^6),
\end{equation}
which corresponds to the second case (B) of Herrera-Varela model
[12] and represents ``electromagnetic mass'' model such that gravitational
mass of a charged fluid sphere originates from the electromagnetic field
alone [1, 18]. This again corresponds to the Tiwari-Rao-Kanakamedala
model [7] with $\alpha = 0 $ case and thus the present model reduces
to isotropic one.

Now, considering the observed values of mass, charge and radius of the electron
(in relativistic units) as $m=6.76\times10^{-56}$ cm, $q=1.38\times
10^{-34}$ cm and $a=10^{-16}$ cm the value of $\alpha$, from the
equation (31), is given by
\begin{equation}
\alpha=-4.77\times10^{95}cm^{-6}.
\end{equation}
For the above value of constant $\alpha$, the energy density in equation (28) becomes
\begin{equation}
\rho(r)= - 6.81 \times 10^{27}(a^2 - r^2) - \frac{\Lambda(r)}{8\pi}.
\end{equation}
The central energy density, ${\rho}_0$, at $r=0$, then can be calculated as
\begin{equation}
{\rho}_0 = - 6.81\times10^{-5} - \frac{\Lambda_0}{8\pi}.
\end{equation}
Thus, from the equation (34) one can see that for $\Lambda_0> 0$ the
energy density of the electron is a negative quantity. It is to be
noted here that in the cosmological context $\Lambda$ positive
is related to the repulsive pressure and hence an acceleration dominated
universe as suggested by the SCP and HZT project report [13, 14,
  19, 20]. However, equation (34) indicates that this negativity
of energy density is also obtainable for $\Lambda_0 < 0$ (which
indicates a collapsing situation of the universe [21]) for its very
small value. In this context it is also possible to
show that at an early epoch of the universe when the numerical value of
negative $\Lambda$ was higher than that of the first term of $\rho$
(i.e. $\sim 10^{-5}$ at $r = 0$) obviously energy density was a positive
quantity. Thus, in the case of decreasing negative value of $\Lambda$
it is clear that there was a smooth crossover from positive energy
density to a negative energy density via a phase of null energy
density! However, these results confirm the vacuum equation of state
$\rho + p_r = 0$ [22 - 25].

We can also see that on the boundary, $r = a$, the total energy density becomes
\begin{equation}
\rho_a = -\frac{\Lambda_a}{8\pi},
\end{equation}
which shows its clear dependency on the cosmological constant. However,
for $\Lambda_a > 0$, $\rho_a$ is  negative whereas for $\Lambda_a < 0$,
$\rho_a$  is as usual a positive quantity. As a simple and interesting
exercise (as all the parameters related to the electron are known) one
can find out the numerical value of $\Lambda_a$, at the boundary of
the spherical system from the equation (35), which equals
$\sim 10^{-7} cm^{-2}$. This constant value of $\Lambda_a$ is too large and
might be related to an early epoch of the universe. Here for finding
out the total energy density $\rho_a$ it is considered that
$\rho_a \leq \rho_{average}$, where $\rho_{average}$ is equal to
$m/\frac{4}{3}\pi a^3$ as the energy density of the spherical
distribution is decreasing from centre to boundary.

\subsection{Model for $\rho + p_r \neq 0$}
 Now using equation (23) the equation (18) can be written as
\begin{equation}
\frac{d}{dr}\left[\tilde{p_r} - \frac{E^2}{8\pi}\right]+
\frac{1}{2}(\tilde{\rho}+ \tilde{p_r}){{\nu}^{\prime}} = \frac{ E^2}{2\pi
r} + 2\alpha q^2r.
\end{equation}
Assuming that the radial stress-energy tensor ${T^1}_{1}=0$, one gets
\begin{equation}
{\nu}^{\prime}= \frac{(e^{\lambda}-1)}{r},
\end{equation}
\begin{equation}
{\tilde p}_r= \frac{E^2}{8\pi}.
\end{equation}
Using equations (37) and (38) in equation (36), we then have
\begin{equation}
\tilde{\rho}+ {\tilde p}_r= \rho + p_r =\frac{(4\alpha q^2 r^2+
E^2/\pi)}{e^\lambda -1}.
\end{equation}
\noindent
Thus, equation (20) takes the form as
\begin{equation}
M = 4\pi\int_{0}^{r}\left[\frac{(4\alpha q^2 r^2 +E^2/\pi)}{e^\lambda
-1}\right]r^2dr.
\end{equation}
\noindent
To make equation (40) integrable we assume that
\begin{equation}
E^2 = \pi k (e^\lambda -1)(1-R^2)-4\pi\alpha q^2 r^2,
\end{equation}
where $k$ is a constant and $R=r/a$, $a$ being the radius of the
sphere.

Thus, the solution set is given by
\begin{equation}
e^{-\lambda} = 1- A R^2 (5 - 3R^2),
\end{equation}
\begin{equation}
e^\nu = (1-2A)^{5/4}e^{\lambda/4}exp[5B{tan^{-1}B(6R^2-5)- \frac{1}{2}tan^{-1}B}],
\end{equation}
\begin{equation}
p_r = \frac{1}{8}k(e^\lambda-1)(1-R^2) - \frac{1}{2}\alpha q^2 r^2 + \frac{\Lambda(r)}{8\pi},
\end{equation}
\begin{equation}
p_{\perp} = \frac{1}{8}k(e^\lambda-1)(1-R^2) + \frac{1}{2}\alpha q^2 r^2 + \frac{\Lambda(r)}{8\pi},
\end{equation}
\begin{equation}
\rho = k(1-R^2)[1 - \frac{1}{8}(e^\lambda -1)] + \frac{1}{2}\alpha q^2 r^2
 - \frac{\Lambda(r)}{8\pi},
\end{equation}
where the constant $A = 8\pi ka^2/15$.

By application of the matching condition at the boundary we again get
the total effective gravitational mass, which in the present case
takes the form
\begin{equation}
m = \frac{8}{15}\pi k a^3 + \frac{q^2}{2a}.
\end{equation}
In view of the equation (41), for vanishing charge the constant $k$
vanishes and hence makes the gravitational mass in the equation (47)
to vanish. Thus, the present case $\rho + p_r = k(1 - R^2) \neq 0$
also represents electromagnetic mass model. \\
Now, the constant $k$ can be expressed in terms of the known values
of the electric mass, radius and charge as
\begin{equation}
k = \frac{15}{16\pi a^4}(2am - q^2).
\end{equation}
At $r=0$, the energy density, from the equation (46), is then given by
\begin{equation}
\rho_0 = - 5.68 \times 10^{-5} - \frac {\Lambda_0}{8\pi}.
\end{equation}
As, in the case of electron, $k$ is a negative quantity so for $\Lambda_0 > 0$ the central
energy density $\rho_0$ is negative only. However, for
$\Lambda_0 < 0$ the central energy density may respectively be
negative and positive depending on the numerical value of $k$ whether it is
higher and lower than that of $\Lambda_0$.

At $r = a$, the total energy density is given by
\begin{equation}
\rho_a = - 4.54 \times 10^{-5} - \frac{\Lambda_a}{8\pi}.
\end{equation}
Similarly, for $\Lambda_a>0$, the energy density is negative whereas for $\Lambda_a<0$,
it may either be negative or positive depending on the numerical value of
$\Lambda_a$ as discussed in the previous case.\\
\subsection{A test model}
 In the previous two cases we have
qualitatively discussed the effect of cosmological parameter
$\Lambda(r)$ on the energy density $\rho(r)$ of the electron.
Let us now explore some quantitative effect and hence treat the
equation (24) in  a different way. If we substitute the value of
$\tilde p_r$, from equation (14), then integrating equation (24) we get
\begin{equation}
\Lambda_{eff} = \Lambda(a) - \Lambda(r) = 8\pi \rho(r) - 8\pi(\alpha {q^2} + \frac{2}{3}\pi\sigma^2_0)(a^2 - r^2),
\end{equation}
where $\Lambda_{eff}$ is the effective cosmological parameter.\\
We study the following cases:\\
For the central value of the energy density of the spherical
distribution, i.e. at $r=0$, the effective cosmological parameter becomes
\begin{equation}
\Lambda_{eff}^0 = \Lambda(a) - \Lambda(0) = 8\pi \rho_0 - 8\pi(\alpha {q^2} + \frac{2}{3}\pi\sigma^2_0)a^2.
\end{equation}
Considering that $\rho_0 \geq \rho_{average}$ the effective
cosmological parameter, at $r = 0$, for the proper numerical values of
the charge and radius of the electron can be found out as
\begin{equation}
\Lambda_{eff}^0 =  1.71 \times 10^{-3} cm^{-2}.
\end{equation}
On the other hand, at the boundary, $r=a$, of the spherical
distribution the effective cosmological parameter becomes
\begin{equation}
\Lambda_{eff}^a =\Lambda(a) - \Lambda(a) = 0.
\end{equation}
Thus, from the equations (53) and (54) it is shown that the effective
cosmological parameter has a finite value at the centre of the
electron which decreases radially and becomes zero at the boundary.\\
\section{Discussions}
We see from the above analysis that the cosmological parameter $\Lambda$ has a definite role even on the energy density of
micro-particle, like electron. We, therefore, feel that it may also be
possible to extrapolate the present investigation to the massive
astrophysical bodies to see the effect of spatially varying
cosmological parameter on their energy densities and vice versa.\\
The proper pressure $p_r$, in general, being positive as
evident from the equation (26) is in accordance with the condition
(21) which may be explained as due to vacuum polarization [26]. In this
connection it is mentioned by Bonnor and Cooperstock [11] that the
negativity of the active gravitational mass and hence negative energy
density for electron of radius $a \sim 10^{-16}$ is consistent with
the Reissner-Nordstr{\"o}m repulsion. We would also like to mention
here that the equation of state in the form $p + \rho = 0$ is
discussed by Gliner [27] in his study of the algebraic
 properties of the energy-momentum tensor of ordinary matter through the
metric tensors and called it the $\rho$-vacuum state of matter. It is also
to be noted that the gravitational effect of the zero-point energies of
particles and electromagnetic fields are real and measurable, as in the
Casimir Effect [28]. According to Peebles and Ratra [29], like all energy,
this zero-point energy has to contribute to the source term in Einstein's
gravitational field equation. This, therefore, demands inclusion of vacuum
energy related term cosmological constant in the field equation. In
this regard it is interesting to recall the comment made by Einstein
[2] where he stated that ``... of the energy constituting matter
three-quarters is to be ascribed to the electromagnetic
field, and one-quarter to the gravitational field'' and did ``disregard'' the
cosmological constant in his field equation is in contradiction to the present
result as shown in the equation (31) and (47). \\

\section*{Acknowledgments}
One of the authors (SR) is thankful to the authority of IUCAA, Pune, India for
providing Associateship programme under which a part of this work was
carried out. Authors thanks are also due to Dr. B. Bhattacharya,
Dr. K. K. Nandi, Dr. R. G. Vishwakarma and the referee for their
valuable comments which made it possible to improve the
article. Special thanks are due to the CTS group, IIT Kharagpur for
constant inspiration throughout the work.

\end{document}